\title{Anisotropy and Physical Viability of General Relativistic Matter}
\documentclass[notitlepage,letterpaper,12pt]{article}
\usepackage[ansinew]{inputenc} 
\usepackage[colorlinks=true,urlcolor=blue,linkcolor=blue,citecolor=blue]{hyperref} 
\usepackage{amsmath}
\usepackage{amsfonts}
\usepackage{amssymb}
\usepackage{graphicx}
\usepackage{subfig}
\usepackage{cancel}
\usepackage{booktabs}
\usepackage{color}
\usepackage{bigints}
\usepackage[top=1cm, bottom=2cm,outer=2cm, inner=2cm]{geometry}
\usepackage{amsthm}
\usepackage{lineno}

\begin{document}
\title {The physical acceptability conditions and  \\ the strategies to obtain anisotropic compact objects}
\author{
\textbf{Daniel Su\'arez-Urango}\thanks{\texttt{danielfsu@hotmail.com}}, \textbf{Laura M. Becerra}\thanks{\texttt{laura.marcela.becerra@gmail.com}} \\
\textit{Escuela de F\'isica, Universidad Industrial de Santander,  }\\ 
\textit{Bucaramanga 680002, Colombia}; \\
\textbf{Justo Ospino}\thanks{\texttt{j.ospino@usal.es}} \\
\textit{Departamento de Matem\'atica Aplicada and} \\
\textit{Instituto Universitario de F\'isica Fundamental y Matem\'aticas,}  \\ 
\textit{Universidad de Salamanca, Salamanca, Spain;} \\
and \textbf{Luis A. N\'{u}\~{n}ez}\thanks{\texttt{lnunez@uis.edu.co}} \\
\textit{Escuela de F\'isica, Universidad Industrial de Santander,}\\ 
\textit{Bucaramanga 680002, Colombia} and \\
\textit{Departamento de F\'{\i}sica,} \\
\textit{Universidad de Los Andes, M\'{e}rida 5101, Venezuela.} 
}

\maketitle

\begin{abstract}
We studied five methods to include anisotropy, or unequal stress distributions, in general relativistic matter configurations. We used nine acceptability conditions that the metric and physical variables must meet to determine if our models were astrophysically viable. Our analysis found the most effective way to introduce anisotropy while keeping a simple density profile. We also found a practical ``rule of thumb'' that relates the density at the boundary to the density at the centre of relativistic matter distributions. Additionally, we calculated the configuration radius and encountered that values observed by NICER for PSR J0740+6620 are consistent with several acceptable matter configurations, both isotropic and anisotropic.
\end{abstract} 
PACS:  \\
Keywords:  
\newpage
\section{Introduction}
\label{Introduction}
General Relativity is experiencing an extraordinary era where what was once considered a mathematical curiosity, such as black holes, and faint phenomena, like gravitational waves, have transformed into observable astrophysical entities~\cite{AbbottEtalLIGOVIRGCol2019, GendreauEtal2022}. Significant efforts explore the properties of physically viable matter configurations that may describe general relativistic compact objects in various states: static, stationary, or undergoing collapse. Any exact solution to the Einstein Equations has certain restrictions, constraining the metric and the energy-momentum tensor to ensure that emerging space-time geometry is astrophysically reasonable. 

Since the seminal result of M.S.R. Delgaty and K. Lake~\cite{DelgatyLake1998}, several works have expanded the set of acceptability conditions to obtain more meaningful solutions~\cite{Ivanov2017, Ivanov2018, HernandezNunezVasquez2018, HernandezSuarezurangoNunez2021, SuarezurangoEtal2022}. These conditions are elaborated under the assumption that there are two distinct components for the pressure, one radial and the other tangential, which yields a richer and more realistic description of the internal structure of a compact object. The consideration of local anisotropy, where the radial and tangential stresses are unequal ($P \neq P_\perp$), has gained recognition as a relevant concept in describing general relativistic stars. This idea can be traced back to the pioneering works of J. H. Jeans~\cite{Jeans1922} and G. Lema\^itre~\cite{Lemaitre1933} and has continued to be explored in both Newtonian and relativistic frameworks (see~\cite{Ruderman1972, BowersLiang1974, CosenzaEtal1981, HerreraNunez1989, HerreraSantos1997, MartinezRojasCuesta2003, HerreraBarreto2004, HerreraEtal2014, Setiawan2019, RahmansyahEtal2020, RahmansyahSulaksono2021, Das2022, KumarBharti2022, RayEtal2023} and references therein). Notably, a recent paper~\cite{Herrera2020} presents intriguing insights into the instability of isotropic pressure distribution in self-gravitating matter systems.

Various heuristic strategies have been employed to describe anisotropic microphysics in astrophysical relativistic matter configurations~\cite{Setiawan2019, RahmansyahEtal2020, RahmansyahSulaksono2021, Sulaksono2015, SetiawanSulaksono2017,  BiswasBose2019}). First, there is the initial method proposed by Bowers \& Liang~\cite{BowersLiang1974}; followed by other schemes such as the proportional-to-gravitation approach~\cite{CosenzaEtal1981}; the quasilocal method~\cite{DonevaYazadjiev2012}; the covariant approach using proportional pressure gradient~\cite{RaposoEtal2019}; the complexity factor method~\cite{Herrera2018}, and the Karmarkar embedding class I~\cite{Karmarkar1948, OspinoNunez2020}. Finally, there is another strategy for implementing anisotropic fluids for General Relativistic matter configurations: the gravitational decoupling approach~\cite{Ovalle2017}.

Throughout this work, we shall consider the first five common assumptions to model non-pascalian fluids in general relativistic matter configurations and examine, through extensive modelling, the consequences of the acceptability conditions. We identify the relevant parameters for a particular equation of state, their range and relevance. We  integrate the structure equations implementing every anisotropic equation of state with the same density profile $\rho(r)$ for all configurations. We also identify a comparable set of parameter ranges so as to compare all the physical acceptabilities of the different anisotropic modelling strategies.

Within this framework, for a particular common density distribution, we explore answers to the following two questions:
\begin{itemize}
    \item Which type of anisotropy strategy leads to more acceptable matter configurations?
    \item Are these acceptable models consistent with the Neutron Star Interior Composition Explorer (NICER) observations~\cite{GendreauEtal2022}? 
\end{itemize}

This paper answers the above questions by organizing our subject matter into several sections. The next section describes the notation and framework of General Relativity. In Section \ref{PhysicalAcceptabilityConditions}, we list the acceptability conditions that our models must meet to be considered candidates for compact stellar objects. Section \ref{AnisotropySec} discusses five approaches to include anisotropy in a general relativistic matter configuration. Next, in Section \ref{Modelling}, we explore the parameter space while fulfilling several acceptability conditions and answer the above queries. Finally, Section \ref{FinalRemarks} summarizes our closing remarks and conclusions.

\section{The field equations}
\label{FieldEquations}
Let us consider the interior of a dense star described by  a spherically symmetric  line element written as
\begin{equation}
\mathrm{d}s^2 = {\rm e}^{2\nu(r)}\,\mathrm{d}t^2-{\rm e}^{2\lambda(r)}\,\mathrm{d}r^2- r^2 \left(\mathrm{d}\theta^2+\sin^2(\theta)\mathrm{d}\phi^2\right),
\label{metricSpherical}
\end{equation}
with regularity conditions at 
$r=r_c=0$, i.e. ${\rm e}^{2\nu_c}=$ constant,  ${\rm e}^{-2\lambda_c}= 1$, and $\nu^{\prime}_c=\lambda^{\prime}_c=0$. 

We shall consider a distribution of matter consisting of a non-Pascalian fluid represented by an energy-momentum tensor:
\begin{equation}
T_\mu^\nu = \mbox{diag}\left[\rho(r),-P(r),-P_\perp(r),-P_\perp(r)  \right] \,,
\label{tmunu}
\end{equation}
where $\rho(r)$ is energy density, with $P(r)$ and $P_\perp(r)$ the radial and  tangential pressures, respectively. 

From Einstein's field equations, we obtain the physical variables in terms of the metric functions as
\begin{eqnarray}
\rho(r)&=& \frac{ {\rm e}^{-2\lambda}\left(2 r \lambda^{\prime}-1\right)+1 }{8\pi r^{2}}\,,\label{FErho} \\
P(r) &=&  \frac {{{\rm e}^{-2\,\lambda}}\left(2r \,\nu^{\prime} +1\right) -1}{8 \pi\,{r}^{2}}\,\label{FEPrad} \qquad \textrm{and} \\
P_\perp(r) &=&-\frac{{\rm e}^{-2\lambda}}{8\pi}\left[ \frac{\lambda^{\prime}-\nu^{\prime}}r-\nu^{\prime \prime }+\nu^{\prime}\lambda^{\prime}-\left(\nu^{\prime}\right)^2\right] \label{FEPtan} \,, 
\end{eqnarray}
where primes $^{\prime}$ denote differentiation with respect to $r$. 

Now,  assuming  the metric function $\lambda(r)$ is expressed in terms of the Misner ``mass''~\cite{MisnerSharp1964} as 
\begin{equation}
 m(r)=\frac{r^2}{2}R^{3}_{232} \; \Leftrightarrow \; m(r)=4\pi \int ^r_0 T^0_0r^2\mathrm{d}r \; \Rightarrow e^{-2\lambda}= 1-\frac{2 m(r)}{r}\, ,
\label{MassDef} 
\end{equation}
Additionally, the interior metric should continuously match the Schwarzschild exterior solution at the sphere's surface,  $r=r_b=R$. This implies that ${\rm e}^{2\nu_b}={\rm e}^{-2\lambda_b}=1-2\mathcal{C}_{\star} = 1 -2M/R$,  where $M = m_b$ is the total mass and $\mathcal{C}_{\star}=M/R$ the compactness of the configuration.  From now on, the subscripts $b$ and $c$ indicate the variable's evaluation at the boundary and the centre of the matter distribution.  
 
The Tolman-Oppenheimer-Volkoff equation (i.e. $T^{\mu}_{r \; ; \mu}~=~0$, the hydrostatic equilibrium equation) for this anisotropic fluid can be written as   
\begin{equation}
\frac{\mathrm{d} P}{\mathrm{d} r} = -\underbrace{(\rho +P)\frac{m + 4 \pi r^{3}P}{r(r-2m)}}_{F_{g}} + \underbrace{\frac{2}{r}\left(P_\perp -P \right)}_{F_{a}} \, .
 \label{TOVStructure1}
\end{equation}
Thus, we can identify two forces competing in compensating the pressure gradient: the ``gravitational force'', $F_{g}$ and the ``anisotropic force'', $F_{a}$. Equation (\ref{TOVStructure1}) together with
\begin{equation}
\label{MassStructure2}
\frac{\mathrm{d} m}{\mathrm{d} r}=4\pi r^2 \rho \,,
\end{equation}
constitute the relativistic stellar structure equations. 

 From equation (\ref{TOVStructure1}), notice that the pressure gradient becomes less steep when the anisotropy is positive $\Delta_{+}~=~P_{\perp}~-~P > 0$, and conversely, it changes more rapidly when the anisotropy is negative $\Delta_{-}~=~P_{\perp}~-~P < 0$. The only possibility for negative anisotropy is that the tangential and radial pressures vanish at $r = r_b$. 
 
 Thus, for a fixed central stiffness,  $\sigma = P_c/\rho_c$, the compactness, $\mathcal{C}_{\star}$, of the sphere increases when there is positive anisotropy $\Delta_{+}$, and decreases when there is negative anisotropy $\Delta_{-}$. Concerning positive anisotropy, we can adjust more massive configurations compared to isotropic $\Delta_{0} = 0$ scenarios. If both forces balance, i.e., $F_{g}=F_{a}$, we obtain a specific matter configuration characterised by vanishing radial pressures and solely sustained by tangential stresses~\cite{Florides1974}. This is because the tangential stresses support the mass shells, reducing the required radial pressure in such circumstances~\cite{HerreraEtal2001}.

\section{The physical acceptability conditions}
\label{PhysicalAcceptabilityConditions}
The emerging physical variables have to comply with the various acceptability conditions~\cite{DelgatyLake1998, Ivanov2017, Ivanov2018, HernandezSuarezurangoNunez2021, SuarezurangoEtal2022}, which are crucial when considering self-gravitating stellar models. Only acceptable self-gravitating objects are of astrophysical interest and, in this work, those models have to comply with nine requirements expressed as~\cite{SuarezurangoEtal2022}:
\begin{enumerate}
\item[{\bf C1:}] $2m/r < 1$, which implies~\cite{Buchdahl1959, Ivanov2002B}:
    \begin{enumerate}
    \item That the metric potentials $\textrm{e}^{\lambda}$ and $\textrm{e}^{\nu}$ are positive, finite and free from singularities within the matter distribution, satisfying $\textrm{e}^{\lambda_{c}} = 1$ and $\textrm{e}^{\nu_{c}}= \mbox{const}$ at the centre of the configuration.
    \item The inner metric functions match the exterior Schwarzschild solution at the boundary surface.
    \item The interior redshift should decrease with increasing of $r$. 
    \end{enumerate}

\item[{\bf C2:}] Positive density and pressures, finite at the centre of the configuration with $P_c=P_{\perp c}$~\cite{Ivanov2002B}.

\item[{\bf C3:}] $\rho^{\prime} < 0$, $P^{\prime} < 0$, $P_{\perp}^{\prime} < 0$ with density and pressures having maximums at the centre, thus $\rho^{\prime}_{c}=P^{\prime}_{c} = P^{\prime}_{\perp c}=0$ with $P_{\perp} \geq P$. 

\item[{\bf C4:}] The causality conditions on the radial,  $0 < v_{s}^2 \leq 1$ and tangential $0 < v_{s \perp}^2 \leq 1$, sound speeds, respectively~\cite{AbreuHernandezNunez2007b}.

\item[{\bf C5:}] The trace energy condition  $\rho - P - 2P_{\perp} \geq 0$, which is more restrictive than the strong energy condition, $\rho + P + 2P_{\perp} \geq 0$, for imperfect fluids~\cite{ Ivanov2018, KolassisSantosTsoubelis1998,PimentelLoraGonzalez2017}. This condition has several interesting consequences for isotropic EoS~\cite{PodkowkaMendesPoisson2018}. 

\item[{\bf C6:}]  The dynamic perturbation analysis restricts the adiabatic index~\cite{HerreraSantos1997, HeintzmannHillebrandt1975, ChanHerreraSantos1993, ChanHerreraSantos1994} 
\[
\Gamma = \frac{\rho + P}{P} v_s^{2} \geq \frac{4}{3} \,.
\]

\item[{\bf C7:}] The Harrison-Zeldovich-Novikov stability condition: $\mathrm{d}M(\rho_c)/\mathrm{d}\rho_c > 0$~\cite{HarrisonThorneWakano1965, ZeldovichNovikov1971}.

\item[{\bf C8:}] The cracking instability against local density perturbations, $\delta \rho = \delta \rho(r)$ (for more details, the reader is referred to~\cite{HernandezNunezVasquez2018, GonzalezNavarroNunez2015, GonzalezNavarroNunez2017}).  

\item[{\bf C9:}] The adiabatic convective stability condition  $\rho^{\prime \prime} \leq 0$, which is more restrictive than the outward decreasing density and pressure profiles~\cite{HernandezNunezVasquez2018}.
\end{enumerate}

Acceptability conditions for general relativistic spheres refer to the criteria that must be satisfied by the metric and physical variables in a relativistic matter distribution to be considered astrophysically viable and consistent within the framework of General Relativity. They are motivated by
\begin{itemize}
    \item \textit{Regularity conditions on the physical and metric variables}, i.e. {\bf C1} and {\bf C2}: A physically acceptable solution should exhibit regular behaviour, particularly at the centre of the sphere, avoiding singularities or divergences in physical quantities such as energy density, pressure, and metric components.
    \item \textit{Energy conditions and equation of state}, i.e. {\bf C2}, {\bf C3}, {\bf C4} and {\bf C5}: Relativistic matter distributions are typically required to satisfy certain energy conditions, which impose constraints on the stress-energy tensor components. These conditions ensure the energy density and pressures associated with the matter distribution are within physically reasonable bounds.
    \item \textit{Stability}, i.e. {\bf C6}, {\bf C7}, {\bf C8} and {\bf C9}:  This involves assessing the stability of the matter distribution against perturbations or dynamic changes, ensuring that it remains in a state of equilibrium and does not collapse, cracks or other undesirable behaviours.
\end{itemize}

\section{Anisotropy heuristic strategies} 
\label{AnisotropySec}
This section will introduce several assumptions and heuristic strategies to model anisotropy in relativistic matter configurations.  Local anisotropy in compact objects is a hypothesis that has gained relevance over time. Nowadays, it is well understood that unequal radial and tangential stresses may increase the stability of neutron star models. However, a complete description of the complex interactions in the fluid that cause such phenomena is still unknown. 

The most common approaches in introducing anisotropy for modelling relativistic matter configuration are:
\begin{itemize}
    \item \textbf{Anisotropy proportional to gravitational force.} M. Cosenza et al. ~\cite{CosenzaEtal1981} inspired by the work of Bowers and Liang~\cite{BowersLiang1974} proposed suitable models for anisotropic matter by considering the anisotropic force proportional to the gravitational one. This relationship leads to the following expression for the difference between the tangential and radial pressures:
    \begin{equation}
    P_{\perp} - P = \frac{C_{GF}\left(\rho + P\right)\left(m + 4\pi r^{3}P\right)}{r - 2m} = \Delta_{GF} \,, \label{DeltaGF}
    \end{equation} 
    \item \textbf{Quasi-local anisotropy} Local anisotropy can also be considered as the influence of quasi-local variables, which are quantities that are not solely dependent on the state of the fluid at a specific point in space-time~\cite{HernandezNunez2004}. These variables, such as the curvature radius $r$ or the compactness $\mu$ ($=2m/r$), are employed as a quasi-local equation of state to describe anisotropy~\cite{DonevaYazadjiev2012}. Within this approach, a particular type of anisotropy is:
    \begin{equation}
    P_{\perp} - P = C_{QL} P \mu = 2 C_{QL} P \frac{ m }{r} = \Delta_{QL} \,. 
    \label{DeltaQL}    
    \end{equation}
    \item \textbf{Anisotropy proportional to a pressure gradient.} Another potential form for the anisotropic force, considering equation (\ref{TOVStructure1}), is for it to be proportional to the pressure gradient. Raposo and collaborators~\cite{RaposoEtal2019} proposed an anisotropy proportional to the covariant derivative of pressure as:
    \begin{equation}
    P_{\perp} - P = -C_{PG} f(\rho)k^{\mu}\nabla_{\mu}P = -C_{PG} f(\rho)\sqrt{1 - \frac{2m}{r}}\frac{\mathrm{d}P}{\mathrm{d}r} = \Delta_{PG} \,, 
    \label{DeltaPG}
    \end{equation}
    where $f(\rho)$ (see appendix \ref{DimensionlessAnisotropies} for details) is a generic function of the energy density and $k^{\mu} = \left(0,k^{1},0,0\right)$ is a unitary space-like vector orthogonal to the fluid four-velocity.
    \item \textbf{Complexity factor anisotropy.} This factor is a quantity defined by decomposing the Riemann tensor, which measures the level of complexity in self-gravitating systems~\cite{Herrera2018, herrera2021, herrera2023}. It reflexes the impact of local anisotropy and density inhomogeneity on the active gravitational mass. Consequently, systems with minimal complexity are represented by homogeneous and isotropic fluids. In the case of anisotropic fluids, satisfying the condition of a vanishing complexity factor with minimal complexity, the local anisotropy can be expressed as follows:
    \begin{equation}
     P_{\perp} - P= - \frac{C_{CF}}{2 r^{3}} \int_{0}^{r} \tilde{r}^{3} \rho^{\prime} \mathrm{d}\tilde{r} = \Delta_{CF}\,. \label{DeltaCF}
     \end{equation}
         \item \textbf{Karmarkar anisotropy.} The Karmarkar condition~\cite{Karmarkar1948} is a relationship among components of the Riemann tensor, given by
    \begin{equation}
    R_{0303}\, R_{1212} -R_{0101}\,R_{2323} -R_{0313}\,R_{0212} = 0\,. \label{KarmarkarCondition1}
    \end{equation}
    This condition provides a geometric mechanism for incorporating anisotropy into matter configurations. To express equation (\ref{KarmarkarCondition1}) in a scalar form, we introduce a set of scalar functions known as structure scalars, obtained from the orthogonal splitting of the Riemann tensor (refer to \cite{OspinoHernandezNunez2017} and \cite{OspinoNunez2020} for more a detailed discussion). Hence, the scalar Karmarkar condition for spherically symmetric static configurations is
    \begin{equation}
        Y_0 X_1+(X_0+X_1)Y_1=0 \,,
    \label{KarmaKcondTetrad}
    \end{equation}
    with
    \begin{equation}
    Y_{0} = 4\pi\left(\rho + 3\mathbf{P}\right),\,\, Y_{1} = \mathcal{E}_{1} - 4\pi\Delta,\,\, X_{0} = 8\pi\rho\quad \text{and} \quad X_{1} = -\left(\mathcal{E}_{1} + 4\pi\Delta\right),
    \end{equation}
    where
    \begin{equation}
    \mathbf{P} = \frac{P + 2P_{\perp}}{3}\quad \textrm{and}\quad \mathcal{E}_{1} = -\frac{4\pi}{r^{3}}\left(\int_{0}^{r} \Tilde{r}^{3}\rho^{\prime}\mathrm{d}\Tilde{r} + r^{3}\Delta\right)\,.
    \end{equation}
    Thus, the induced anisotropy by the Karmarkar condition, written in terms of the physical variables, is given by
    \begin{equation}
    P_{\perp} - P = \frac{ C_{KC}}{r^{3}} \int_{0}^{r} \Tilde{r}^{3}\rho^{\prime}\mathrm{d}\Tilde{r} \, \left(\frac{(3P -\rho) - \frac{1}{r^{3}} \int_{0}^{r} \Tilde{r}^{3}\rho^{\prime}\mathrm{d}\Tilde{r} }{4 \rho}\right) = \Delta_{KC}\,.
    \label{DeltaKC}
    \end{equation}
\end{itemize}
In the above equations (\ref{DeltaGF}), (\ref{DeltaQL}), (\ref{DeltaPG}), (\ref{DeltaCF}), and (\ref{DeltaKC}), we denoted the corresponding anisotropic parameter by $C_{GF}$, $C_{QL}$, $C_{PG}$, $C_{CF}$ and $C_{KC}$, respectively.

The relationship between the complexity and Karmarkar anisotropies is evident  when we re-write equation (\ref{DeltaKC}) in terms of $\Delta_{CF}$, i.e.
     \begin{equation}
     \label{KarmarkarComplexity}
      \Delta_{KC} = -\Delta_{CF}\left( \frac{C\left(3P-\rho\right) + 2\Delta_{CF}}{2 \rho C} \right) \, .  
     \end{equation}
Where we have set $C = C_{CF} = C_{KC}$. It is worth mentioning that when $\Delta_{CF} = 0 \quad \Leftrightarrow \quad \Delta_{KC} =0$ and the only matter configuration for both anisotropic strategies corresponds to the Schwarzchild homogeneous isotropic solution. 

Another strategy for implementing anisotropic fluids in General Relativistic matter configurations is the gravitational decoupling approach~\cite{Ovalle2017}. This procedure assumes that the energy-momentum tensor splits into two parts as
     \begin{equation} 
     T_{\mu}^{\nu} = \hat{T}_{\mu}^{\nu} + \theta_{\mu}^{\nu}\,,
     \label{tmunuGD}
     \end{equation}
where $\hat{T}_{\mu}^{\nu}$ corresponds to the perfect fluid contribution and $\theta_{\mu}^{\nu}$ describes any other coupled form of gravitational source. Implementing the anisotropic parameter $C$ for modelling this method is unattainable. Thus, comparing the models emerging from this strategy with those executed with all previous techniques is impossible. It deserves a more detailed consideration which will be developed elsewhere.

\section{Anisotropy and physical acceptability}
\label{Modelling}
In this section, we discuss the physical acceptability of relativistic anisotropic models. We numerically integrate the structure equations (\ref{TOVStructure1}) and (\ref{MassStructure2}) implementing every equation of state for anisotropy from the previous section, (i.e. $\Delta_{GF}$, $\Delta_{QL}$, $\Delta_{PG}$, $\Delta_{CF}$,  and $\Delta_{KC}$) and selecting a common density profile 
\begin{equation}
    \rho\left(r\right) = \rho_{c}\left(1 - \alpha r^{2}\right)\,, \label{densityprofile}
\end{equation}
where the central density $\rho_{c}$ and the constant $\alpha$ are free parameters. 

This simple Tolman VII density profile~\cite{Tolman1939} is not deprived of physical interest~\cite{RaghoonundunHobill2015} and has a long tradition of modelling compact objects. It corresponds to the Gokhroo-Mehra~\cite{GokhrooMehra1994} solution used in several anisotropic static spheres in General Relativity~\cite{HernandezNunez2004, Stewart1982, HerreraEtal2001}. Additionally, under some circumstances~\cite{Martinez1996}, it leads to densities and pressures that give rise to an equation of state similar to the Bethe-B\"orner-Sato Newtonian equation for nuclear matter~\cite{HohlerEtal1973}. It also describes radiating anisotropic fluid spheres~\cite{Martinez1996, HernandezNunezPercoco1999} representing the Kelvin-Helmholtz phase in the birth of a neutron star~\cite{HerreraMartinez1998A, HerreraMartinez1998B}.

Now, from equation (\ref{densityprofile}), we obtain the boundary radius of the configuration as a function of the physical parameters of the problem, i.e. $\rho_{c}$ and $\varkappa =  \rho_b/\rho_c$ as 
\begin{equation}
    \tilde{\rho} = \tilde{\rho}_{c}\left(1-\tilde{\alpha} x^{2}\right) \quad \Rightarrow  \tilde{m} = 4\pi\tilde{\rho}_{c}\left(\frac{x^{3}}{3} - \Tilde{\alpha}\frac{x^{5}}{5}\right) \quad \Rightarrow R = \left\{\frac{ M }{4\pi\rho_{c}\left[\frac{1}{3}-\left(1-\varkappa\right)\frac{1}{5}\right]}\right\}^{1/3} \,.
    \label{ConfigurationRadious}
\end{equation}
Where we have defined the following quantities
\begin{align}
   \Tilde{\alpha} =\alpha R^{2}\,, \quad  \varkappa = 1 - \Tilde{\alpha} =\frac{\rho_{b}}{\rho_{c}}\,, \quad   m = R\tilde{m}\,, \quad \rho = \frac{1}{R^{2}}\tilde{\rho}\,, \quad \text{and} \quad r = R x\,, \label{DimentionlessVariables}
\end{align}
where $R$ and $M$ are the structure boundary radius and total mass, respectively. In appendix \ref{DimensionlessAnisotropies}, we present the dimensionless expressions for the structure equations and  each anisotropic EoS, including some information about the simple numerical integration techniques.

\subsection{The range of the parameters}
The solution of equation (\ref{TOVStructure1}) for the density profile (\ref{densityprofile}) is sensitive to  $\rho_{c}$, $\varkappa =  \rho_b/\rho_c$ and the anisotropy factor, $C$~\cite{RaghoonundunHobill2015}. A variation of these three factors generates a parameter space, exhibiting several acceptability conditions satisfied by each model. We shall identify which anisotropy delivers more physically acceptable configurations, i.e. satisfy more acceptability conditions. Thus, we shall identify a common set of parameter variations so as to compare the physical acceptability of the different anisotropic modelling strategies.  

We start determining the possible variation of a common anisotropic parameter, $C$. Regarding the case of $\Delta_{GF}$, observe that equation (\ref{TOVStructure1}) leads to
\begin{equation}
    \frac{\mathrm{d}P}{\mathrm{d}r} = - h \frac{(\rho + P)(m + 4 \pi  r^3 P)}{r(r-2m)}\,,
\label{ansatzcosenza}
\end{equation}
with  $h = 1 - 2 C$, and when $h = 1$ the isotropic case is recovered.  Notice that condition {\bf C3} and equation (\ref{ansatzcosenza}) implies $h > 0$, therefore if $\rho_{b} \neq 0$ we have, 
\begin{equation}
h = 1 - 2 C > 0 \,\, \Rightarrow \,\, C < \frac{1}{2} \,, \quad \text{and since }\quad  P_{\perp} \geq 0 \,\, \Rightarrow \,\, 0 \leq C < \frac{1}{2} \, .
\label{PtGreaterP}
\end{equation}
The tangential pressure should be positive at the boundary $P_{\perp \, b} \geq 0$ within the matter distribution and from equation (\ref{DeltaGF}), it restricts the anisotropic parameter to  $0 \leq C < \frac{1}{2}$ for any EoS having $\rho_{b} \neq 0$.  We selected six values for the anisotropy parameter, i.e. $C = 0.000, \, 0.050, \, 0.150, \, 0.250, \, 0.350,$ and $0.450$. 

In addition to the anisotropic parameter, $C$, there are two other significant elements: the central density, $\rho_{c}$, and  $\varkappa$. According to typical values in compact objects/neutron stars,  $\rho_{c}$ could go from $0.1 \times 10^{15}$ to $2.5 \times 10^{15}$ ${\rm g/cm^{3}}$. The scale variation for $\varkappa$ runs from $0.0$ (vanishing density at the surface) to $0.9$ (almost homogeneous density). Finally, we have to provide the total mass, $M$, of the configuration ($\approx 2.08\,M_{\odot}$, the highest reliable gravitational mass of any neutron star~\cite{ MillerEtal2021, RileyEtal2021}) to determine the central density of each model.

\subsection{The best method to introduce anisotropy}
To answer the first question, we shall follow two lines of reasoning in identifying which of the above anisotropy strategies is best suited in providing more acceptable models. The next section identifies regions in the parameter space ($C$, $\varkappa$ and $\rho_c$), that comply with the acceptability criteria. Figure \ref{AceptaAnisotropies} displays, in a colour scale, those patches for five different values of the anisotropy factor $C$.  For example, in the isotropic case, i.e. $C = 0$, we obtain $33$ of these physically fully acceptable models. More acceptable matter configurations are placed below the red line in all cases shown, i.e. when 
\begin{equation}
\label{rhoboundary}
\rho_b \leq \frac{9}{10} \rho_c \left( 1 - \frac{2 \rho_{c}}{5}\right) \, .     
\end{equation}
As displayed in Figure \ref{Deltas}, the second approach is to sum up the total number of models satisfying all nine-acceptability criteria. This method, discussed in section \ref{TotalNumberAcceptableModels}, complements the previous criterion because we explore the number of possible acceptable models for the whole range of variation of the anisotropic parameter.   

\subsubsection{Acceptable model distribution in a parameter space}
\label{parameterspace}
\begin{figure}[t!]
\centering
\includegraphics[width=6.5in]{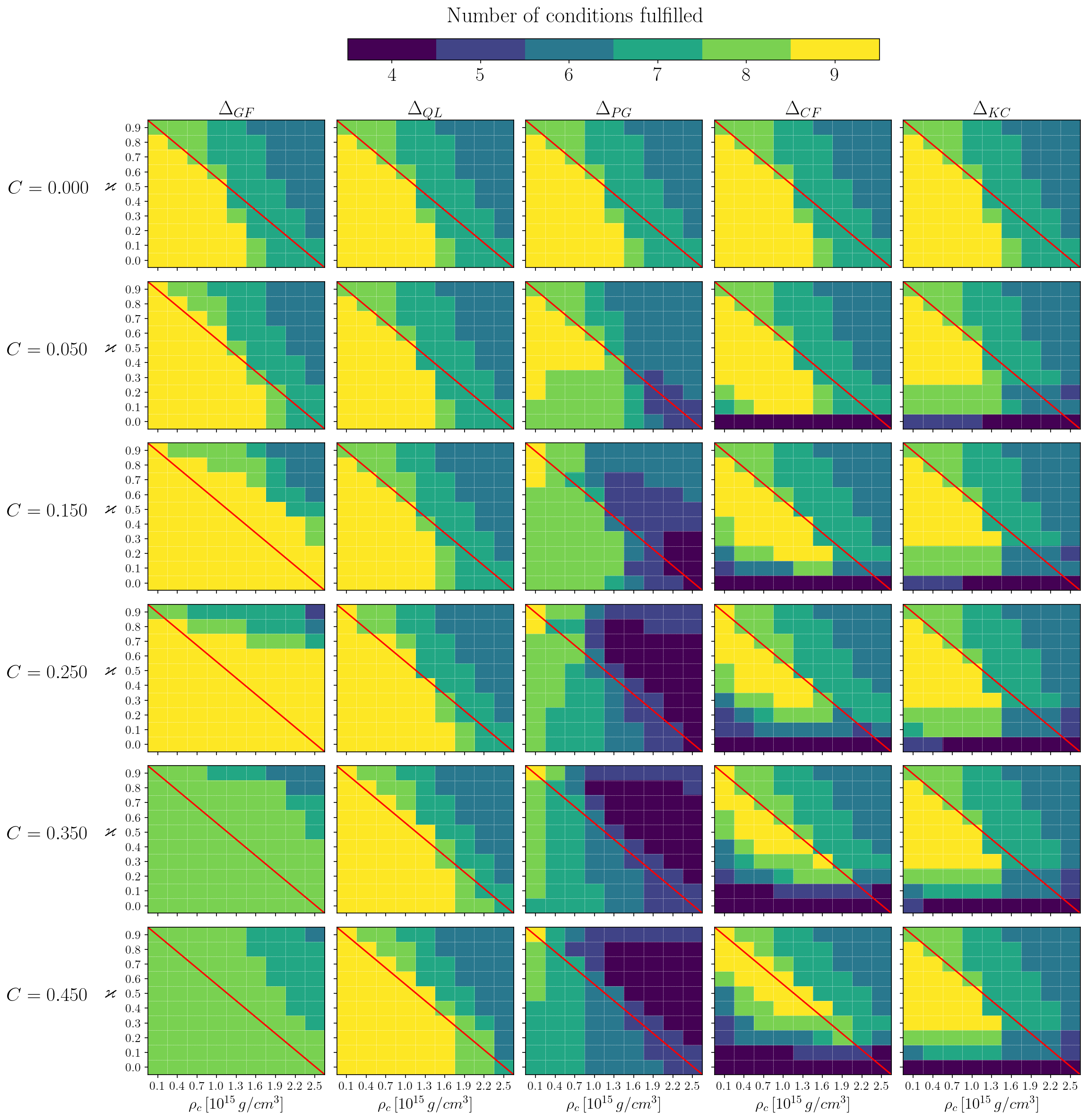}
\caption{Number of conditions fulfilled by the different anisotropy heuristic strategies. The strategies $\Delta_{GF}$ and $\Delta_{QL}$ lead to better modelling (more yellow squares representing models complying with nine criteria) than the isotropic case for $C = 0.050, 0.150$ and $0.250$, This is not the case for the other anisotropic approaches: $\Delta_{CF}$, $\Delta_{KC}$, which have the maximum number of acceptable models in the isotropic condition. When the anisotropic parameter increases to $C = 0.150$, the number of acceptable models boost for $\Delta_{GF}$ and very little for $\Delta_{QL}$ and $\Delta_{KC}$. For the other approaches, the number of models decreases. Increasing the anisotropy to $C = 0.250$ generates a maximum of acceptable models with the $\Delta_{GF}$ method. When $C=0.350$, no $\Delta_{GF}$-model satisfies the nine acceptability criteria. The $\Delta_{QL}$-strategy gives a few new models, and other anisotropic methods decrease the number of acceptable configurations. Nearly all matter distributions meet the nine criteria for the considered values of $\varkappa$ and $\rho_c$. Finally, when $C=0.450$, only the $\Delta_{QL}$-strategy provides more acceptable models than the isotropic condition. Clearly, for all the anisotropic methods, we found more acceptable models when $\rho_b \leq \frac{9}{10} \rho_c \left( 1 - \frac{2 \rho_{c}}{5}\right)$.}
\label{AceptaAnisotropies}
\end{figure}

In this section, we shall discuss the acceptable model distribution in the common parameter space defined by $0.000 \leq C \leq 0.450$; $\, 0.1 \times 10^{15}\,\leq  \rho_c \leq 2.5 \times 10^{15}\, {\rm g/cm^{3}}$ and $0.0 \leq \varkappa \leq 0.9$. Figure \ref{AceptaAnisotropies} displays, in a colour scale, this model distribution for six different values of the anisotropy factor $C$. As will be clear in the following discussion, this range of variation in the parameter plane $(\rho_c, \varkappa)$ is due to the NICER-acceptable models~\cite{MillerEtal2021} when we considered the total mass of the configuration $M \approx 2.08 M_{\odot}$. See Figure \ref{Radios208} to grasp the rationale of the parameter variation. 

For a low anisotropic presence ($C = 0.050$ displayed in Figure \ref{AceptaAnisotropies}), $\Delta_{GF}$ and $\Delta_{QL}$ strategies deliver more acceptable models (yellow patches represent models satisfying the nine criteria) than in the isotropic case. Several models with anisotropy proportional to the pressure gradient, $\Delta_{PG}$, become unacceptable because they do not meet the adiabatic index's stability criterium \textbf{C6}. On the other hand, configurations with anisotropy defined by the complexity factor have nonphysical negative pressure and positive tangential pressure gradient. Moreover, configurations with vanishing density at the boundary ($\rho_b = 0 \rightleftharpoons \varkappa = 0.0$) do not comply with \textbf{C5}, \textbf{C6} or \textbf{C8}. The Karmarkar anisotropy scheme, $\Delta_{KC}$, produces unsuitable configurations having negative tangential pressures. When $\varkappa = 0.0$ the corresponding matter distributions do not comply with \textbf{C4}, \textbf{C6} or \textbf{C8}. In general, as $\varkappa$ increases, the speed of sound exceeds the light speed, failing \textbf{C4}. The increase in the central density oversteps the condition on the trace of the energy-momentum tensor, and the darker region in the upper right corner is due to the models' cracking (\textbf{C8}).

\begin{figure}[t!]
\centering
\includegraphics[width=6in]{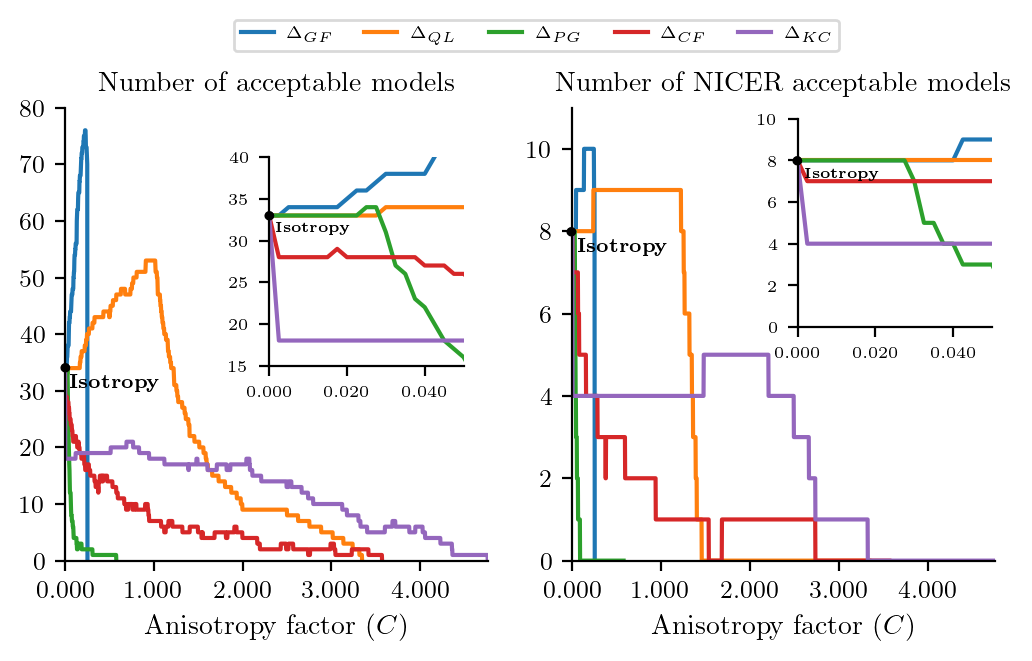}
\caption{In the left panel, we present the number of acceptable models $N(C)$ fulfilling the nine acceptability conditions as a function of the anisotropic factor $C$ for each anisotropic strategy. The embedded plot displays a zoom of the region near the isotropy, i.e. $C \approx 0$. The black point represents the number of nine-condition-acceptable-isotropic models. There are more anisotropic-acceptable models than isotropic ones. $\Delta_{GF}$ has a limited range of anisotropy variation, $0.000 \leq C_{GF} \leq 0.250$, with a pronounced peak for 76 models for $C_{GF} = 0.225$. The number of acceptable quasi-local models, $\Delta_{QL}$, increases with anisotropy, reaching 53 for $C_{QL} = 0.910$. The areas under the curves are proportional to the total number of anisotropic configurations described by the three parameters: $C$, $\rho_c$ and $\varkappa$.  The quasi-local anisotropic modelling, $\Delta_{QL}$, is the most effective anisotropic strategy,  followed by the Karmarkar scheme, $\Delta_{KC}$, and then by the complexity factor approach, $\Delta_{CF}$. The right panel illustrates the number of acceptable models representing the NICER observational estimation of the radius $R$ for PSR J0740+6620. The area under the curve for the anisotropic strategies follows the same pattern in the NICER-acceptable models. More possible models exist for the $\Delta_{QL}$ approach.}
\label{Deltas}
\end{figure}

Raising $C$ to $0.150$ enhances the acceptability when anisotropy is proportional to the gravitational force, and $65$ out of $90$ models satisfy all the conditions. On the other hand, quasi-local anisotropy does not show variation in the acceptable configurations, while the number of acceptable models with $\Delta_{PG}$ decreases drastically by failing with condition \textbf{C6}. The increased anisotropy in $\Delta_{CF}$ makes acceptable models, now to have positive tangential pressure gradient and negative radial pressure. One model becomes acceptable for $\Delta_{KC}$ by fulfilling condition \textbf{C5}.

Increasing the anisotropic factor $C$ to $0.250$ further causes a maximum in $\Delta_{GF}$. As seen from Figure \ref{AceptaAnisotropies}, almost all models, $\Delta_{GF}$, meet the nine criteria for the considered values of $\varkappa$ and $\rho_c$. This is also evident from Figure \ref{Deltas}. It attains $76$ fully acceptable models when $C=0.225$. The remaining 12 configurations mainly do not satisfy the causality condition of radial sound speed \textbf{C4}.  Quasi-local anisotropy also increases, to a lesser extent, the number of acceptable models. Only one model for $\Delta_{PG}$ becomes unacceptable due to condition \textbf{C6}. Regarding $\Delta_{CF}$, a few models are no longer acceptable, breaking the tangential pressure condition \textbf{C3}. Acceptable models with $\Delta_{KC}$ remain unchanged. 

As it is clear from Figure \ref{AceptaAnisotropies}, when $C=0.350$, there is no $\Delta_{GF}$-model satisfying the nine acceptability criteria; the $\Delta_{QL}$-strategy allows a few new models, while $\Delta_{PG}$ and $\Delta_{CF}$ decrease in acceptable models. In contrast, $\Delta_{KC}$ remains unchanged. Finally, as displayed in Figure \ref{AceptaAnisotropies}, i.e. for $C=0.450$, only the $\Delta_{QL}$-strategy provides more acceptable models than the isotropic condition. Several models under the red line, for $\Delta_{CF}$, become acceptable by fulfilling condition \textbf{C4}.

\begin{table}[ht]
    \centering
    \caption{Values of the anisotropic parameter $C$ and the number of acceptable models $N(C)$ in the range of the considered $\varkappa$ and $\rho_c$. $C_0$ represents the anisotropic parameter where no other fully acceptable model is found. $C_{max}$ represents the central value among the $C$ values that yield the highest number of acceptable models for a particular anisotropy strategy. $N(C_{max})$, is greater than the number of acceptable isotropic models for  $\Delta_{GF}$, $\Delta_{QL}$ and $\Delta_{PG}$ strategies. The area under the curve, $N_{Total} \propto \int_{0}^{C_0}{\rm d}C \; N(C) $, for each anisotropic strategy. Clearly, for the range in $\varkappa$ and $\rho_c$ considered, there are more anisotropic acceptable models than their isotropic counterparts. Concerning the NICER-acceptable models, we found $N_{NICER}(C_{NICER-max}) = 10,  9, 8, 8$ and   $8$, for $\Delta_{GF}$, $\Delta_{QL}$, $\Delta_{PG}$, $\Delta_{CF}$ and $\Delta_{KC}$, respectively and the area under the curve for the anisotropic strategies follow the same pattern. Figure \ref{tab:varkappa} illustrates these results.}    
\begin{tabular}{|l|r|r|r|r|r|} \hline
\textbf{Anisotropy \& Number of acceptable Models} & $\Delta_{GF}$& $\Delta_{QL}$& $\Delta_{PG}$ & $\Delta_{CF}$ & $\Delta_{KC}$ \\ \hline
   $C_0$                      & 0.253        & 3.345        & 0.578      & 3.570        & 4.760         \\ \hline
   $C_{max}$                  & 0.225        & 0.910        & 0.025      & 0.000        & 0.000         \\ \hline
   $N(C_{max})$               & 76           & 53           & 34         & 33           & 33        \\ \hline     
   $N_{Total} \propto area =\int_{0}^{C_0}{\rm d}C \; N(C) $  
                              & 14.236       & 76.524       & 2.649      & 22.799       & 57.846        \\ \hline   
   $C_{NICER-0}$              & 0.253        & 1.458        & 0.088      & 1.538        &  3.328        \\ \hline 
   $C_{NICER-max}$            & 0.188        & 0.730        & 0.014      & 0.000        &  0.000        \\ \hline
   $N_{NICER}(C_{NICER-max})$ & 10           & 9            & 8          & 8            &  8        \\ \hline     
   $N_{NICER-Total} \propto area =\int_{0}^{C_0}{\rm d}C \; N_{NICER}(C) $  
                              & 2.332       & 11.775        & 0.370      & 4.695        & 11.955        \\ \hline   
\end{tabular}
    \label{tab:CandN}
\end{table}

\subsubsection{Total number of acceptable models and the best anisotropy strategy}
\label{TotalNumberAcceptableModels}
This section extends our analysis by determining the total number of entirely acceptable models. To achieve this, we redefine the range for the anisotropic parameter, starting from isotropy ($C=0.000$) and continuing until the last value ($C=C_0$) where no other fully acceptable model exists, i.e. satisfying all nine criteria for acceptability.

In Figure \ref{Deltas} and Table \ref{tab:CandN}, we show the most effective methods for introducing anisotropy. The most useful approach is the quasi-local $\Delta_{QL}$ method~\cite{DonevaYazadjiev2012}, followed by the Karmarkar scheme~\cite{Karmarkar1948, OspinoNunez2020}, and finally the complexity factor approach~\cite{Herrera2018, herrera2021, herrera2023}. These strategies include varying degrees of effectiveness due to the significant range of the anisotropic parameter $C$: $0.000 \leq C_{QL} \leq 3.345$, $0.000 \leq C_{KC} \leq 4.760$, and $0.000 \leq C_{CF} \leq 3.570$, respectively.

Anisotropy leads to more acceptable models than isotropic ones. The most distinct
scheme is the anisotropy proportional to gravitational force, $\Delta_{GF}$. It has a narrow range of variation for the anisotropic parameter, with $0.000 \leq C_{GF} \leq 0.250$, and a pronounced peak with 76 models for $C_{GF} = 0.225$. Around half of the models fall within $0.050 \leq C_{GF} \leq 0.250$. Models with $C_{GF} > 0.250$ are considered unacceptable due to their positive tangential pressure gradient, which violates condition \textbf{C3}. Regarding the quasi-local approach, $\Delta_{QL}$, the number of models satisfying all requirements increases with the level of anisotropy, reaching a peak of 53 for $C_{QL} = 0.910$. This method has a significant range in anisotropy $0.000 \leq C_{QL} \leq 3.345$. The simplest but least effective method is  anisotropy proportional to pressure, $\Delta_{PG}$. It has a limited anisotropic range of $0.000 \leq C_{PG} \leq 0.578$, with a maximum of 34 models at very low anisotropy, $C_{PG} = 0.025$. The following method is complexity anisotropy, $\Delta_{CF}$, which has a considerable range, $0.000 \leq C_{CF} \leq 3.570$, but no anisotropic parameter generates more acceptable models than the isotropic case. Finally, the geometric Karmarkar strategy is $\Delta_{KC}$ related to the complexity anisotropy and has the broadest range, i.e., $0.000 \leq C_{CF} \leq 4.760$.

\begin{table}[t!]
    \centering
    \caption{Range of $\varkappa$ and $\rho_c$ for the PSR J0740+6620 with $2.08~\pm~0.07~M_{\odot}$, considering distinct anisotropy approaches. In this table, we present the range of $\varkappa$ for the possible values of $\rho_c$ consistent with the observational NICER data. We found various possible configurations for different $\rho_c$, ranging from almost homogeneous density profiles, i.e. $\varkappa \approx 0.8$, to others with vanishing density at the boundary of the configuration where $\varkappa \approx 0.0$. Assuming a simple density profile, the observed radius for PSR J0740+6620, can be described by acceptable isotropic matter configurations and several anisotropic approaches.  Models with the highest $\varkappa$ correspond to anisotropic configurations with $\Delta_{GF}$ strategy. }    
    \begin{tabular}{|c||c|c|c|} \hline
    Range of $\varkappa$& $\rho_c = 1.3 \times 10^{15}~{\rm g/cm}^3$ &  $\rho_c = 1.0 \times 10^{15}~{\rm g/cm}^3$ & $\rho_c = 0.7 \times 10^{15}~{\rm g/cm}^3$ \\ \hline \hline
       $C_{GF} = 0.225$& $0.0 \leq \varkappa \leq 0.1$          & $      0.1 \leq \varkappa \leq 0.3$          & $      0.4 \leq \varkappa \leq 0.8$ \\ \hline
       $C_{QL} = 0.910$& $0.0 \leq \varkappa \leq 0.1$          & $      0.1 \leq \varkappa \leq 0.3$          & $      0.4 \leq \varkappa \leq 0.7$ \\ \hline
       $C_{PG} = 0.025$& $0.0 \leq \varkappa \leq 0.1$          & $      0.1 \leq \varkappa \leq 0.3$          & $      0.4 \leq \varkappa \leq 0.6$ \\ \hline
       $C_{iso} = 0.000$& $0.0 \leq \varkappa \leq 0.1$          & $      0.1 \leq \varkappa \leq 0.3$          & $      0.4 \leq \varkappa \leq 0.6$ \\ \hline
    \end{tabular}
    \label{tab:varkappa}
\end{table}

\subsection{NICER acceptable models}
\label{NICERAcceptable}
The Neutron Star Interior Composition Explorer is an X-ray telescope on the International Space Station which studies the X-ray emissions from neutron stars, helping to determine their size, mass, and the properties of their dense interiors. By measuring the mass and radius of multiple neutron stars, NICER refines our understanding of the equation of state, providing valuable constraints on the properties of ultra-dense matter. NICER also detects the pulsation of neutron stars, permitting scientists to explore the dynamics of their atmospheres, unravelling the physical processes occurring in and around them. NICER has been employed to obtain the first precise (and dependable) measurements of a pulsar's size and mass and the first-ever map of hot spots on its surface (see~\cite{GendreauEtal2022} and references therein). 

\begin{figure}[t!]
\centering
\includegraphics[width=6in]{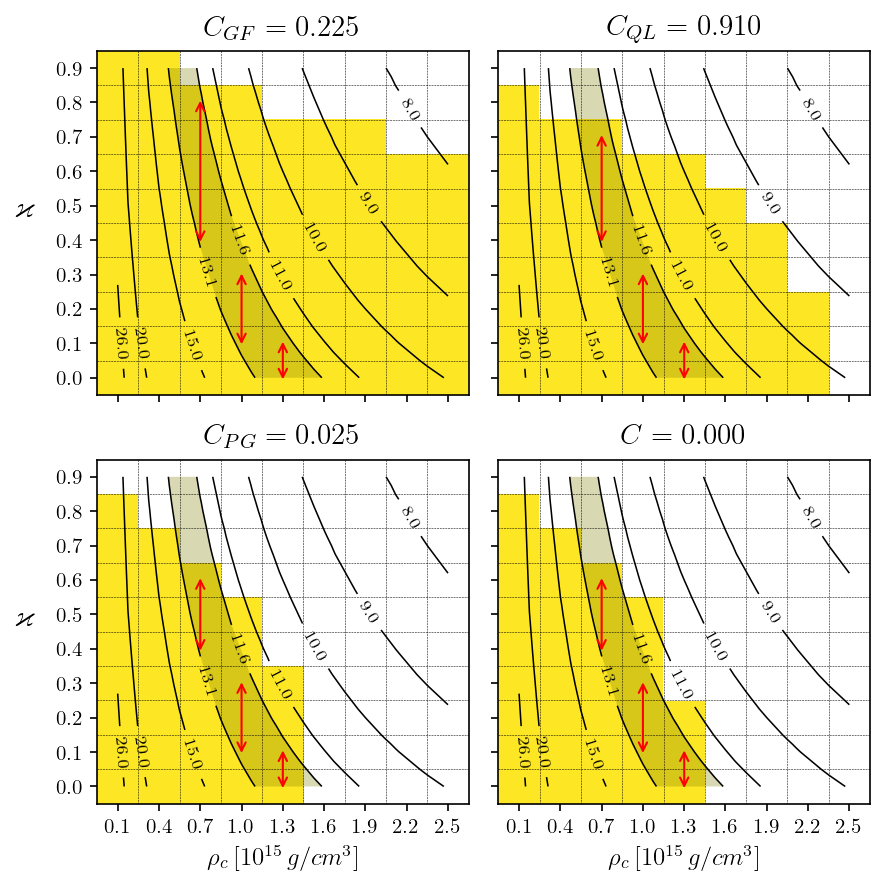}
\caption{NICER acceptable models for PSR J0740+6620. There could be several descriptions (isotropic and anisotropic) for this pulsar. We consider three anisotropic methods, $\Delta_{GF}$, $\Delta_{QL}$, $\Delta_{PG}$ and the isotropic case. We present the isotropic scenario because the number of acceptable models is the same for the $\Delta_{CF}$ and $\Delta_{KC}$. We show that the value of all possible radii ranges within $11.6~{\rm km}~\leq~R~\leq~13.1~{\rm km}$, corresponding to several NICER-acceptable models with the selected parameter space, fulfilling all the physically acceptable conditions. The NICER acceptable region overlaps with more yellow models in $\Delta_{GF}$ and $\Delta_{QL}$ plots than in the other anisotropic strategies. The red arrows indicate the possible values for the central density, $0.7~\times~10^{15}~{\rm g/cm}^3~\leq~\rho_c~\leq~1.3~\times~10^{15}~{\rm g/cm}^3$ and the corresponding values of $\varkappa$ consistent with the assumed density profile (\ref{densityprofile}), which are displayed in Table \ref{tab:varkappa}.   }
\label{Radios208}
\end{figure}

In Figure \ref{Radios208}, we include a region covering observational estimates for the radius of the PSR J0740+6620, with mass $2.08~\pm~0.07~M_{\odot}$, which is the highest reliable gravitational mass of any neutron star~\cite{ MillerEtal2021, RileyEtal2021}. In Table \ref{tab:CandN}, we indicate the number of NICER-compatible models for different anisotropic strategies.  

From equation (\ref{ConfigurationRadious}), we calculate the configuration radius as a function of $\rho_c$ and $\varkappa$. We find that the central density, $0.7~\times~10^{15}~{\rm g/cm}^3~\leq~\rho_c~\leq~1.3~\times~10^{15}~{\rm g/cm}^3$, is consistent with observational data. The range for $\varkappa$ associated with various central densities for different anisotropy strategies are displayed in Table~\ref{tab:varkappa}.

We also find various acceptable configurations with different $\rho_c$, ranging from almost homogeneous density profiles, i.e. $\varkappa \approx 0.8$, to others with vanishing density at the configuration boundary where $\varkappa \approx 0.0$. Assuming a simple density  (\ref{densityprofile}), the observed radius for PSR J0740+6620 can be described by acceptable isotropic matter configurations and several anisotropic approaches. Models with the highest $\varkappa$ correspond only to anisotropic configurations within $\Delta_{GF}$ strategy.

Figure \ref{Radios208} provides a visual representation of the results detailed in Table \ref{tab:CandN}.  All possible radii $11.6~{\rm km}~\leq~R~\leq~13.1~{\rm km}$ harmonise with various NICER-acceptable models. The red arrows denote workable values for the central density, ranging from $0.7~\times~10^{15}~{\rm g/cm}^3$ to $1.3~\times~10^{15}~{\rm g/cm}^3$, coupled with their corresponding $\varkappa$ values that align with the assumed density profile (\ref{densityprofile}).


In both the $\Delta_{GF}$ and $\Delta_{QL}$ anisotropy strategies, the NICER acceptable region encompasses more yellow models compared to other anisotropic approaches. Specifically, there are ten NICER models for $\Delta_{GF}$ with $C_{GF}=0.225$  and nine for $\Delta_{QL}$ with $C_{QL}=0.910$. However, if we consider the whole range of possible values for the anisotropy parameter $C$,  the total number of NICER models reaches approximately $N_{NICER-Total-QL} \approx 11.775$ for the $\Delta_{QL}$ strategy (see Table~\ref{tab:CandN}).

\section{Final remarks}
\label{FinalRemarks}
This work introduces the most common assumptions in modelling non-pascalian fluids in general relativistic matter configurations. Local anisotropy in compact objects is a hypothesis that has gained relevance over time. So far, however, it is still not well known how unequal radial and tangential stresses may increase the stability of neutron star models. The complete description of the complex interactions in the fluid that cause such phenomena is unknown~\cite{HerreraSantos1997, KumarBharti2022}.

We explore five different heuristic methods to include anisotropy in general relativistic matter configurations. We found that the most effective approach in introducing anisotropy, with a physically meaningful density profile (\ref{densityprofile}) is the quasi-local $\Delta_{QL}$ method~\cite{DonevaYazadjiev2012}; followed by the Karmarkar scheme~\cite{Karmarkar1948, OspinoNunez2020}; and last by the complexity factor approach~\cite{Herrera2018, herrera2021, herrera2023}. Incorporating any of the five types of anisotropy schemes considered in this study results in a significantly greater number of acceptable configurations than their isotropic counterparts within the specified range of critical parameters ($C$, $\varkappa$ and $\rho_c$).

Furthermore, as shown in Figure \ref{AceptaAnisotropies} and from equation (\ref{rhoboundary}), we have established a ``rule of thumb'' that provides a simple relationship between the density at the boundary, $\rho_b$, and the centre, $\rho_c$, for relativistic matter distributions. This rule can serve as a helpful tool for identifying potentially realistic and acceptable models of compact objects. By leveraging this relationship, researchers can make informed judgments about the physical viability of different matter configurations.

From equation (\ref{ConfigurationRadious}), we calculate the configuration radius as a function of $\rho_c$ and $\varkappa$. We found that the central density, $0.7\times10^{15}~{\rm g/cm}^3~\leq~\rho_c~\leq~1.3\times10^{15}~{\rm g/cm}^3$, is consistent with NICER-observational data. In Table~\ref{tab:varkappa}, we introduce the corresponding ranges for $\varkappa$. All the possible radii values, $11.6~{\rm km}~\leq~R~\leq~13.1~{\rm km}$, correspond to several NICER-acceptable models within the selected parameter space, fulfilling all physically acceptable conditions.

Assuming a simple density profile (\ref{densityprofile}), the observed radius for PSR J0740+6620 can be described by acceptable isotropic matter configurations and several anisotropic approaches. Models with the highest $\varkappa$ correspond only to anisotropic configurations with $\Delta_{GF}$ strategy.

\section*{Acknowledgments}
L.A.N. acknowledges  the financial support of the Vicerrector\'ia de Investigaci\'on y Extensi\'on de la Universidad Industrial de  Santander and Universidad de Salamanca through the  research mobility programs. L.A.N. also thanks the hospitality of the Departamento de Matem\'aticas Aplicadas, Universidad de Salamanca. The Vicerrector\'ia de Investigaci\'on y Extensi\'on, Universidad Industrial de Santander Postdoctoral Fellowship Program No. 2023000359 supported L.M.B.  J.O.  acknowledges financial support from Ministerio de Ciencia, Innovaci\'on y Universidades (grant PGC2018-096038-B-100) and Junta de Castilla y Le\'on  (grant SA083P17). D.S.U and L.A.N gratefully thank the internship program of the ERASMUS+ project, Latin-American alliance for Capacity buildiNG in Advance physics (LA-CoNGA physics), where this paper's first ideas and calculations began.  D.S.U. thanks for the hospitality of the Departamento de F\'isica of the Colegio de Ciencias e Ingenier\'ia, Universidad San Francisco de Quito and especially to Dr Ernesto Contreras for the fruitful discussions.

\section*{Appendices}
\appendix
\section{The structure equations}
We determine the physical variables ($\rho, m, P, P_{\perp}$) and check the acceptability conditions. We compare the physical acceptability among models with the same parameters ($\rho_{c}$, $\alpha$, $C$) having different anisotropy strategies. The more acceptable models an anisotropy generates, the more it may represent observable compact objects.
Now, expressing the structure equations (\ref{TOVStructure1}) and (\ref{MassStructure2}) in term of dimensionless quantities  we have
\begin{align}
    \frac{\mathrm{d}\tilde{P}}{\mathrm{d}x} &= -\frac{\left(\tilde{\rho} + \tilde{P}\right)\left(\tilde{m} + 4\pi\tilde{P}x^{3}\right)}{x\left(x-2\tilde{m}\right)} + \frac{2 \left(\tilde{P}_{\perp} - \tilde{P}\right)}{x} \label{dimensionlessTOV} \qquad \text{and} \\
    \frac{\mathrm{d}\tilde{m}}{\mathrm{d}x} &= 4\pi\tilde{\rho}x^{2} \,, \label{dimensionlessmassgradient}
\end{align}
with
\begin{equation}
    \tilde{\rho} = \tilde{\rho}_{c}\left(1-\tilde{\alpha} x^{2}\right) \quad \Rightarrow \quad  \tilde{m} = 4\pi\tilde{\rho}_{c}\left(\frac{x^{3}}{3} - \Tilde{\alpha}\frac{x^{5}}{5}\right)\,,
    \label{dimensionlessdensity}
\end{equation}
leaving only equation (\ref{dimensionlessTOV}) to be integrated. We have this new set of dimensionless physical variables:
\begin{align}
    m = R\tilde{m}\,, \quad P = \frac{1}{R^{2}}\tilde{P}\,, \quad P_{\perp} = \frac{1}{R^{2}}\tilde{P}_{\perp}\,, \quad \rho = \frac{1}{R^{2}}\tilde{\rho}\,, \quad \text{and} \quad r = R x\,, \label{ChangeofVariables}
\end{align}
where $R$ is the boundary radius of the configuration. 

It is convenient to transform the parameter $\Tilde{\alpha}$ ($=\alpha R^{2}$) into a quantity with greater physical meaning. Evaluating the dimensionless density in (\ref{dimensionlessdensity}) at the surface of the configuration, where $x=1$, we can define
\begin{equation}
    \varkappa = 1 - \Tilde{\alpha} = \rho_{b}/\rho_{c}
\end{equation}
as the density ratio at the surface to the density at the centre.

\section{Dimensionless equations of state for anisotropy} \label{DimensionlessAnisotropies}
The change of variables proposed in (\ref{ChangeofVariables}) to express equations in dimensionless form has the virtue of preserving the equations without additional constants. That is, we can directly put the tilde mark on the variables (and swap $r$ for $x$ in the case of the radial coordinate) to obtain the dimensionless version. Equations (\ref{dimensionlessTOV}) and (\ref{dimensionlessmassgradient}) are an example of what has just been stated. However, here are some simple calculations that prove it.

\begin{itemize}
    \item \textbf{Anisotropy proportional to gravitational force.}Carrying out the change of variables (\ref{ChangeofVariables}) on the anisotropy proportional to gravitational force yields
    \begin{equation*}
    \frac{\tilde{P}_{\perp} - \tilde{P}}{R^{2}} = \frac{C_{GF}\left(\frac{\tilde{\rho}}{R^{2}} + \frac{\tilde{P}}{R^{2}}\right)\left(R\tilde{m} + 4\pi R^{3}x^{3}\frac{\tilde{P}}{R^{2}}\right)}{Rx - 2R\tilde{m}}\,.
    \end{equation*}
    Now, rearranging the constant $R$ to the right-hand side, we have
    \begin{equation}
    \tilde{P}_{\perp} - \tilde{P} = \frac{C_{GF}R^{2}\left(\frac{\tilde{\rho}}{R^{2}} + \frac{\tilde{P}}{R^{2}}\right)R\left(\tilde{m} + 4\pi R^{2}x^{3}\frac{\tilde{P}}{R^{2}}\right)}{R\left(x - 2\tilde{m}\right)}\,,
    \end{equation}
    from where we have that
    \begin{equation}
    \tilde{\Delta}_{GF} = \frac{C_{GF}\left(\tilde{\rho} + \tilde{P}\right)\left(\tilde{m} + 4\pi x^{3}\tilde{P}\right)}{x - 2\tilde{m}}\,.
    \end{equation}
    \item \textbf{Quasi-local anisotropy.} Quasi-local anisotropy is a more straightforward case since compactness is a dimensionless variable. Implementing the change of variables (\ref{ChangeofVariables}) in equation (\ref{DeltaQL}) leads us to
    \begin{equation*}
    \frac{\tilde{\Delta}_{QL}}{R^{2}} = \frac{2C_{QL}R\tilde{m}\frac{\tilde{P}}{R^{2}}}{Rx}\,,
    \end{equation*}
    and therefore
    \begin{equation}
    \tilde{\Delta}_{QL} = \frac{2 C_{QL} \tilde{m}\tilde{P}}{x}\,.
    \end{equation}
    \item \textbf{Anisotropy proportional to a pressure gradient.} In this particular case, we first choose the function $f\left(\rho\right) = \rho$ as in \cite{RaposoEtal2019}, leaving anisotropy (\ref{DeltaPG}) as
    \begin{equation*}
    \Delta_{PG} = -C_{PG}\rho\sqrt{1 - \frac{2m}{r}}\frac{\mathrm{d}P}{\mathrm{d}r}\,.
    \end{equation*}
    Therefore, the anisotropy factor $C_{PG}$ has dimensions length cubed. Thus, substituting the change of variables into the last equation gives
    \begin{equation*}
    \frac{\tilde{\Delta}_{PG}}{R^{2}} = -R^{3}\tilde{C}_{3}\frac{\tilde{\rho}}{R^{2}}\sqrt{1 - \frac{2R\tilde{m}}{Rx}}\frac{1}{R^{3}}\frac{\mathrm{d}\tilde{P}}{\mathrm{d}x}\,,
    \end{equation*}
    and consequently, we get
    \begin{equation}
    \tilde{\Delta}_{PG} = -\tilde{C}_{PG}\tilde{\rho}\sqrt{1 - \frac{2\tilde{m}}{x}}\frac{\mathrm{d}\tilde{P}}{\mathrm{d}x}\,.
    \end{equation}
    \item \textbf{Complexity factor anisotropy} We can solve the integral in the anisotropy (\ref{DeltaCF}) since density profile is given, yielding
    \begin{equation*}
    \Delta_{CF} = \frac{\rho_{c}\alpha r^{2}}{5}\,.
    \end{equation*}
    Now, applying the change of variables, we get
    \begin{equation*}
    \frac{\tilde{\Delta}_{CF}}{R^{2}} = \frac{1}{5}\frac{\tilde{\rho}_{c}}{R^{2}}\frac{\tilde{\alpha}}{R^{2}}R^{2}x^{2}\,,
    \end{equation*}
    and therefore
    \begin{equation}
    \tilde{\Delta}_{CF} = \frac{\tilde{\rho}_{c} \left(1-\varkappa\right) x^{2}}{5}\,.
    \end{equation}
    \item \textbf{Karmarkar anisotropy.} Given the density profile (\ref{densityprofile}), we can compute derivatives and integrals to obtain
    \begin{equation}
    \Delta_{KC  } = \frac{\rho_{c}\alpha r^{2}}{5\rho}\left(\frac{\rho - 3P}{2} - \frac{\rho_{c}\alpha r^{2}}{5}\right)\,.
    \end{equation}
    Therefore the dimensionless induced anisotropy by the Karmarkar condition is given by
    \begin{equation}
    \tilde{\Delta}_{KC} = \frac{\tilde{\rho}_{c} \left(1-\varkappa\right) x^{2}}{5\tilde{\rho}}\left(\frac{\tilde{\rho} - 3\tilde{P}}{2} - \frac{\tilde{\rho}_{c} \left(1-\varkappa\right) x^{2}}{5}\right)\,.
    \end{equation}

\end{itemize}

\section{Numerical integration}
\label{NumericalAppendix}
Equation (\ref{dimensionlessTOV}) was numerically integrated with Python, implementing the \textit{RK45} method through the \textit{solve\_ivp} function. The solution was started at the surface of the model, with initial values $x_{b} = 1$ and $\Tilde{P}_{b} = 0$, and proceeded with an adaptive step towards the centre, with final values $x_{c} = 10^{-15}$ and $P\left(x_{c}\right) = P_{c}$. Since $x$ takes values between $10^{-15}$ and $1$ we can identify $R$ as the total radius.


\end{document}